\documentclass[pas,manuscript,showpacs,showkeys,superscriptaddress,nofootinbib]{revtex4-1}
\usepackage{graphicx}
\usepackage{epsfig}  
\usepackage{epsf}    
\usepackage[tbtags]{amsmath}
\usepackage{amsfonts}
\usepackage{float}
\usepackage{subfig}
\usepackage{dcolumn}
\usepackage{bm}
\usepackage{dcolumn}
\usepackage{textcomp}
\usepackage{multirow}
\usepackage{slashed}
\usepackage{float}
\restylefloat{figure}
\usepackage[]{hyperref}
  \hypersetup{
  bookmarks=true,         
  unicode=true,         
  pdftoolbar=true,     
  pdfmenubar=true,     
  pdffitwindow=true,    
  pdfstartview={FitH},    
  pdfsubject={Sterilenus@DUNE},  
  pdfnewwindow=true,     
  pdfcreator={RevTeX},
  colorlinks=true,     
  linkcolor=red,   
  citecolor=blue,    
  filecolor=black,  
  urlcolor=blue,           
  }
\usepackage{hypcap}

\newcommand{\beq}{\begin{equation}}
\newcommand{\eeq}{\end{equation}}
\newcommand{\beqa}{\begin{eqnarray}}
\newcommand{\eeqa}{\end{eqnarray}}

\newcommand\barparen[1]{\overset{(-)}{#1}}

\begin{document}
\title{Constraining general U(1) interactions from neutrino-electron scattering measurements at DUNE near detector}
\author{Kaustav Chakraborty}
\email[Email Address: ]{kaustav@prl.res.in}
\affiliation{Physical Research Laboratory, Navrangpura, Ahmedabad-380009, India}

\author{Arindam Das}
\email[Email Address: ]{adas@particle.sci.hokudai.ac.jp}
\affiliation{Institute for the Advancement of Higher Education, Hokkaido University, Sapporo 060-0817, Japan}
\affiliation{Department of Physics, Hokkaido University, Sapporo 060-0810, Japan}

\author{Srubabati Goswami}
\email[Email Address: ]{sruba@prl.res.in}
\affiliation{Physical Research Laboratory, Navrangpura, Ahmedabad-380009, India}

\author{Samiran Roy}
\email[Email Address: ]{samiran@prl.res.in}
\affiliation{Physical Research Laboratory, Navrangpura, Ahmedabad-380009, India}

\preprint{EPHOU-21-015}

\begin{abstract}
{The neutrino-electron scattering process is a powerful tool to explore new physics beyond the standard model. Recently the possibility of DUNE Near Detector (ND) to constrain various new physics scenarios using this process have been highlighted in the literature. In this work, we consider the most general U(1) model and probe the constraints on the mass and coupling strength of the additional $Z'$ from $\nu-e$ scattering at DUNE ND. The presence of the $Z^\prime$ gives rise to extra interference effects. In the context of the general U(1) model, the destructive interference can occur in either neutrino or anti-neutrino channel or for both or none. This opens up the possibilities of getting four different type of signal in the neutrino and ant-neutrino runs of DUNE. We perform the analysis using both the total rate and binned events spectrum. Our results show that in a bin by bin analysis the effect of destructive interference is less compared to the analysis using total rate. We present the bounds on the $ m_{Z^\prime} - g_X$ plane from $\nu-e$ scattering measurements at DUNE ND and compare these with those obtained from other $\nu-e$ scattering, COHERENT, and beam dump experiments. We show that the DUNE ND can give the best bound for certain mass ranges of $Z^\prime$. }
\end{abstract}


\maketitle


\section{Introduction}
\label{sec1}
The Standard Model (SM) of particle physics is remarkably successful in explaining  almost all the phenomena observed in nature. 
 However, it fails to account for the small neutrino masses as is required by the observation of neutrino oscillation in several terrestrial experiments.
It  also does not provide any explanation  for the existence of  Dark Matter  in the universe. 
Other indications to a beyond SM picture includes the observed matter-antimatter asymmetries of the universe,  existence of dark energy, the recent results of flavour anomalies etc.  

The path to the new physics is not very clear at this moment and 
various extensions of the SM have been considered in the literature.
The most  economical  renormalizable extension of the SM is to augment 
it with an  extra U(1) gauge group. 
Such U(1) extensions can arise in the context of string inspired models and    Grand Unified Theories 
with rank higher than four, such that the symmetry group can break into  $G_{SM} \times U(1)^n$ with $n > 1$ \cite{Langacker:2008yv,HEWETT1989193}. 
 A general U$(1)$ extension of the SM  includes three singlet Right Handed Neutrinos (RHNs) to cancel the gauge and mixed gauge-gravity anomalies. 
 After the breaking of the general U$(1)$ symmetry, the Majorana mass term of the RHNs is generated which induces the seesaw mechanism to generate the tiny neutrino mass.    
  Such an extension also involves a neutral and beyond the standard model (BSM) gauge boson, $Z^\prime$, which acquires mass  after the U$(1)$ breaking which in turn needs a singlet scalar boson. 

  A common and interesting U$(1)$ extension is the B$-$L model \cite{PhysRevD.20.776,Marshak:1979fm,Davidson:1980br,Mohapatra:1980qe,Wetterich:1981bx,Masiero:1982fi} which is a special case of the general U(1) scenario. For B$-$L case, the left handed and right handed fermions are equally charged  under the U$(1)$ gauge group. However, in a general U$(1)$ scenario, the left handed and right handed fermions are differently charged \cite{Appelquist:2002mw,Das:2017flq,Das:2021esm}. In such a case, the left handed and right handed fermions  couple differently to $Z^\prime$. Hence the effect of the general U(1) charges are manifested in a different way in the interaction between fermions and $Z'$ as compared to B-L picture.  
 One can also have the flavour non-universal models where the anomaly cancellation occurs  within each family and one can have family dependent U(1) symmetries \cite{Foot:1990mn} like 
 $L_i - L_j$ with $ i,j = e,~\mu,~\tau$.   

 The bound on  the mass and interaction strength of an additional  $Z^{\prime}$   in the context of U(1) models have been  studied  extensively in the literature. 
 For $Z^{\prime}$ mass  around  Electroweak scale/TeV  scale  the constraints can come from  collider 
searches  \cite{Dittmar:2003ir,Basso:2008iv,Das:2016zue,Accomando:2017qcs,Ekstedt:2016wyi,Das:2019pua}, the  most popular channel being the  dilepton channel \cite{ATLAS:2016bps,CMS:2016cfx,ATLAS:2019erb} 
and from elctroweak precision data  \cite{Erler:2009jh}. 
The current experimental bounds  from LEP and ATLAS  and CMS detectors of  the Large Hadron Collider are summarized in \cite{Zyla:2020zbs}.    
 Possibilities of probing a lower mass  $Z^\prime$, assuming it does not couple directly to the SM particles have been explored in the context of LHC 
 in \cite{Abdallah:2021npg}.  
 Lower mass of $Z^\prime$ ($m_{Z'} \lesssim 10$ GeV) with interaction strength  lower than $10^{-2}$ can be constrained from various  experiments 
 like  neutrino-electron scattering \cite{Bilmis:2015lja,Sevda:2017wwn} and beam-dump experiments \cite{PhysRevLett.57.659,PhysRevLett.67.2942,Blumlein:2013cua,Alekhin:2015byh,FASER:2019aik}. Constraint on very low coupling strength ($\lesssim 10^{-7}$) and low mass region can come from SN1987A \cite{Dent:2012mx,Kazanas:2014mca}.

 Different  general U$(1)$ scenarios that are relevant for solving the flavour problem in the context of two Higgs doublet model \cite{Campos:2017dgc} 
 have been considered in \cite{Lindner:2018kjo} and constraints were obtained from  TEXONO \cite{TEXONO:2009knm}, CHARM-II \cite{VILAIN1993351} and GEMMA \cite{Beda:2009kx} data. In recent times it has been realized that  the upcoming  high precision neutrino oscillation experiments can also provide a powerful testing ground to explore physics beyond the SM. Specially the potential of the proposed DUNE Near Detector (ND) \cite{DUNE:2020ypp} to probe non-oscillation new physics has been well studied in literature \cite{Kelly:2020dda,Ballett:2019bgd,Bakhti:2018avv,Dev:2021qjj,Breitbach:2021gvv,DeRomeri:2019kic,Coloma:2020lgy}. In particular, the prospect of the neutrino-electron scattering process at DUNE have been highlighted for instance in \cite{Bischer:2018zcz,deGouvea:2019wav}.
This process provides a clean channel for precision measurements in SM as well as BSM scenarios \cite{deGouvea:2006hfo}. 
  In this context, the constraints on $Z^\prime$ interaction for  
 Leptophilic models via neutrino-electron scattering at DUNE have been obtained in \cite{Ballett:2019xoj}.   
 More recently  in \cite{Dev:2021xzd}  the  U(1)$_{B-L}$ and $L_\mu - L_e$ models  have been constrained from neutrino-electron scattering at DUNE ND.

 In this paper we consider the most general U$(1)$ scenarios and the possibility of probing this via neutrino-electron scattering at the DUNE ND. We obtain the U$(1)$ charges of the fermions from the cancellation of the gauge and gravitational anomalies in terms of the two free parameters. Assuming different representative values of these parameters, the constraints are derived on the mass and coupling strength of $Z^\prime$ employing a bin by bin analysis of $\nu -e $ scattering at DUNE ND.  
 
We compare our results with that obtained in U(1)$_{B-L}$ and leptophilic $L_\mu - L_e$ model and point out the salient features of the  different U$(1)$ scenarios. We also highlight the differences between total rate only \cite{Dev:2021xzd} and bin by bin analysis. Further, we include the constraints obtained from
other electron scattering experiments like TEXONO, CHARM-II, BOREXINO \cite{Bellini:2011rx}, BABAR \cite{BaBar:2014zli}, Orsay \cite{DAVIER1989150}, E141 \cite{Andreas:2012mt,Bjorken:2009mm} and delineate the parameter space where the DUNE $\nu$-e scattering data gives the best constraints.  
 
 The paper is organized as follows : in the next section we briefly summarize the model and present the neutrino-electron scattering cross sections and discuss the special features due to general U(1) charges. The relevant details of the experiments considered in our analysis have been presented in section \ref{Exp Details} followed by detailed analysis and results obtained in section \ref{results_DUNE}. Finally, we draw the conclusion in section \ref{Conclusion}.


\section{Neutrino-electron scattering in U(1) extended Model} We investigate a general U$(1)$ extension of the SM governed by the gauge group $SU(3)_{c}\times SU(2)_{L} \times U(1)_{Y} \times U(1)_{X}$. 
It includes an SM singlet scalar field $(\Phi)$ along with the SM Higgs doublet $(H)$. The extra singlet scalar is responsible for breaking U$(1)_X$ symmetry. The cancellation of all the gauge and the mixed gauge-gravity anomalies in this scenario necessitates the inclusion of three SM singlet RHNs.  
In Tab.~\ref{tab1} we depict the particle content of the model and the corresponding charge assignments. The U$(1)_X$ charges of the particles can be expressed in terms $x_H$ (U(1)$_{X}$ charge of Higgs doublet) and $x_{\Phi}$ (U$(1)_X $ charge of singlet scalar) \cite{Das:2021esm}. These charges are obtained from the following anomaly cancellation conditions:

\begin{table}[h]
\begin{center}
\begin{tabular}{||c|ccc||rcr||}
\hline
\hline
            & SU(3)$_c$ & SU(2)$_L$ & U(1)$_Y$ & \multicolumn{3}{c||}{U(1)$_X$}\\[2pt]
\hline
\hline
&&&&&&\\[-12pt]
$q_L^i$    & {\bf 3}   & {\bf 2}& $\frac{1}{6}$ & $x_q^\prime$ 		& = & $\frac{1}{6}x_H + \frac{1}{3}x_\Phi$  \\[2pt] 
$u_R^i$    & {\bf 3} & {\bf 1}& $\frac{2}{3}$ & $x_u^\prime$ 		& = & $\frac{2}{3}x_H + \frac{1}{3}x_\Phi$  \\[2pt] 
$d_R^i$    & {\bf 3} & {\bf 1}& $-\frac{1}{3}$ & $x_d^\prime$ 		& = & $-\frac{1}{3}x_H + \frac{1}{3}x_\Phi$  \\[2pt] 
\hline
\hline
&&&&&&\\[-12pt]
$\ell_L^i$    & {\bf 1} & {\bf 2}& $-\frac{1}{2}$ & $x_\ell^\prime$ 	& = & $- \frac{1}{2}x_H - x_\Phi$   \\[2pt] 
$e_R^i$   & {\bf 1} & {\bf 1}& $-1$   & $x_e^\prime$ 		& = & $- x_H - x_\Phi$  \\[2pt] 
\hline
\hline
$N_R^i$   & {\bf 1} & {\bf 1}& $0$   & $x_\nu^\prime$ 	& = & $- x_\Phi$ \\[2pt] 
\hline
\hline
&&&&&&\\[-12pt]
$H$         & {\bf 1} & {\bf 2}& $-\frac{1}{2}$  &  $-\frac{x_H}{2}$ 	& = & $-\frac{x_H}{2}$\hspace*{12.5mm}  \\ 
$\Phi$      & {\bf 1} & {\bf 1}& $0$  &  $2 x_\Phi$ 	& = & $2 x_\Phi$  \\ 
\hline
\hline
\end{tabular}
\end{center}
\caption{The particle content of the general U(1)$_X$scenario where i
is the family indices for the three generations. See text for details.}
\label{tab1}
\end{table}
\begin{align}
{\rm U}(1)_X \otimes \left[ {\rm SU}(3)_c \right]^2&\ :&
			2x_q^\prime - x_u^\prime - x_d^\prime &\ =\  0, \nonumber \\
{\rm U}(1)_X \otimes \left[ {\rm SU}(2)_L \right]^2&\ :&
			3x_q^\prime + x_\ell^\prime &\ =\  0, \nonumber \\
{\rm U}(1)_X \otimes \left[ {\rm U}(1)_Y \right]^2&\ :&
			x_q^\prime - 8x_u^\prime - 2x_d^\prime + 3x_\ell^\prime - 6x_e^\prime &\ =\  0, \nonumber \\
\left[ {\rm U}(1)_X \right]^2 \otimes {\rm U}(1)_Y&\ :&
			{x_q^\prime}^2 - {2x_u^\prime}^2 + {x_d^\prime}^2 - {x_\ell^\prime}^2 + {x_e^\prime}^2 &\ =\  0, \nonumber \\
\left[ {\rm U}(1)_X \right]^3&\ :&
			{6x_q^\prime}^3 - {3x_u^\prime}^3 - {3x_d^\prime}^3 + {2x_\ell^\prime}^3 - {x_\nu^\prime}^3 - {x_e^\prime}^3 &\ =\  0, \nonumber \\
{\rm U}(1)_X \otimes \left[ {\rm grav.} \right]^2&\ :&
			6x_q^\prime - 3x_u^\prime - 3x_d^\prime + 2x_\ell^\prime - x_\nu^\prime - x_e^\prime &\ =\  0. 
\label{anom-f}
\end{align}
The Yukawa interactions in this model can be written as 
\begin{equation}
{\cal L}^{\rm Yukawa} = - Y_u^{\alpha \beta} \overline{q_L^\alpha} H u_R^\beta
                                - Y_d^{\alpha \beta} \overline{q_L^\alpha} \tilde{H} d_R^\beta
				 - Y_e^{\alpha \beta} \overline{\ell_L^\alpha} \tilde{H} e_R^\beta
				- Y_\nu^{\alpha \beta} \overline{\ell_L^\alpha} H N_R^\beta- Y_N^\alpha \Phi \overline{N_R^{\alpha c}} N_R^\alpha + {\rm h.c.},
\label{LYk}
\end{equation}
where $\tilde{H} \equiv i  \tau^2 H^*$. The neutrino mass is generated from the last two terms of Eq.~\ref{LYk} after symmetry breaking. The requirement of the Yukawa Lagrangian to respect the U$(1)_X$ symmetry gives the following conditions:
\begin{eqnarray}
-\frac{x_H}{2} 		&=& - x_q^\prime + x_u^\prime \ =\  x_q^\prime - x_d^\prime \ =\  x_\ell^\prime - x_e^\prime=\  - x_\ell^\prime + x_\nu^\prime \, \nonumber \\
2 x_\Phi	&=& - 2 x_\nu^\prime. 
\label{Yuk}
\end{eqnarray} 

Solving the sets of equations \ref{anom-f} and \ref{Yuk} allow us to obtain the individual U$(1)_X$ charges of the fermions in the model in terms of $x_H$ and $x_\Phi$. 
Interestingly, we notice that the U$(1)_X$ charge of the left and right handed components of the fermions are different unlike the U$(1)_{B-L}$ scenario which corresponds to $x_H=0$ and $x_{\Phi}=1$. 

After the breaking of the U$(1)_X$ symmetry, the mass of the new gauge boson ($Z^\prime$) is generated as $m_{Z^\prime}=\frac{g_X}{2}\sqrt{(4 v_\Phi x_\Phi)^2+ (x_H v_h)^2}$ \cite{Das:2021esm}
where $v_\Phi$ and $v_h$ are the U$(1)_X$ and SM vacuum expectation values (VEV) respectively. Here $g_X$ is the U$(1)_X$ gauge coupling. The existence of such a neutral BSM gauge boson will allow additional interactions with the fermions : 
\begin{equation}
-\mathcal{L}_{\rm{int}}^{f}= g_X (\overline{\ell_L} Q_X^\ell \gamma^\mu Z^\prime_\mu {\ell_{L}} + \overline{\ell_R} Q_X^{e_R} \gamma^\mu Z^\prime_\mu \ell_R ) + g_X (\overline{q_L} q_X^\ell \gamma^\mu Z^\prime_\mu {q_{L}} + \overline{q_R} q^{u(/d)}_{X_R} \gamma^\mu Z^\prime_\mu q_R )
\label{Lint}
\end{equation}
where $Q_X^\ell = -\frac{1}{2}x_H-x_\Phi$, $Q_X^{e_R}= -x_H-x_\Phi$ and $q_X^\ell = \frac{1}{6}x_H+ \frac{1}{3} x_\Phi$, $q^{u}_{X_R}= \frac{2}{3}x_H+ \frac{1}{3} x_\Phi $, $q^{d}_{X_R}= -\frac{1}{3}x_H+ \frac{1}{3} x_\Phi $.
 

\begin{figure}[h]
\centering
\includegraphics[width=17cm]{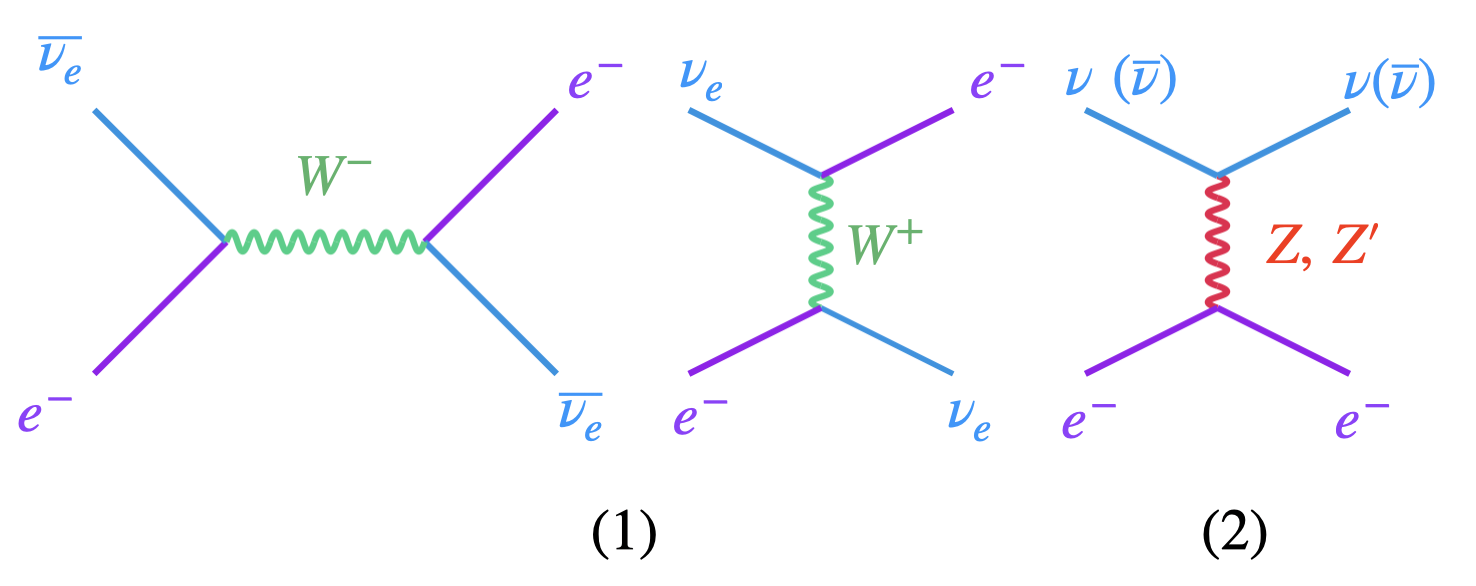}
\caption{The electron-neutrino scattering by the charged (1) and neutral (2) mediators in a general U$(1)_X$ scenario. The $Z^\prime$ vertices manifest the U$(1)_X$ charges.}
\label{fig1}
\end{figure}

The interaction between the light neutrinos and the electrons through the light $Z^\prime$ will explicitly show the effect of the general U$(1)_X$ charges. Several cases are of interest:

(i) The most popular special case is U(1)$_{B-L}$ which corresponds to $x_H=0$ and $x_{\Phi}=1$. This implies $Q_X^\ell =Q_X^{e_R} $. Therefore, the left and right handed fermions couple to $Z'$ with equal strength. 

(ii) If $x_{\Phi}=-x_{H}$ then $Q_X^{e_R}=0$. An example of this scenario with $x_{H}=-1$ and $x_{\Phi}=1$ will be studied in the subsequent sections.
 
(iii) If $x_{\Phi}=-1/2 \, x_{H}$ then $Q_X^\ell=0$. This case is not of relevance for our studies as the neutrinos do not couple to the electrons.
 
 (iv) The most general case corresponds to $Q_X^\ell \neq Q_X^{e_R} $ implying left and right handed leptons couple differently to $Z'$ unlike U(1)$_{B-L}$ leading to interesting consequences.
 
The Fig.~\ref{fig1} shows the Feynman diagrams for the charged and neutral current mediated $\nu-e$ scattering processes in a general U$(1)_X$ model. For the sake of completeness of U(1) scenario, we also consider the leptophilic model $L_{\mu} - L_e$
in our analysis. Following the scattering processes shown in Fig.~\ref{fig1}, we estimate the complete differential scattering cross section $\dfrac{d \sigma(\nu e)}{dT}$( with respect to the recoil kinetic energy $(T)$ of the outgoing electron ) including the interference effects. The SM cross section for $\nu -e$ scattering mediated by the $W$ and $Z$ bosons is given by
\begin{equation}
\dfrac{d \sigma (\nu e)}{dT}\bigg\vert_{\rm{_{_{SM}}}} = \dfrac{2G^2_{F} m_e}{\pi E^2_{\nu}}(a^2_1 E^2_{\nu} +a^2_2(E_{\nu}-T)^2 - a_1 \, a_2 \, m_e T),
\label{SM-1}
\end{equation}
where $E_{\nu}$ is the energy of the incoming neutrino, $G_{F}$ is the Fermi constant, $m_e$ is the mass of electron, and $T \, (0 < T <\dfrac{2E^2_{\nu}}{2E_{\nu} + m_e} )$ is the recoil kinetic energy of the outgoing electron. The values of $a_1$ and $a_2$ for various flavor of neutrinos (anti-neutrinos) are given in Table. \ref{tab2}.
\begin{table}[htb]
\small
 \begin{tabular}{|c|c|c|} 
 \hline
 Scattering Process  & $a_1$ & $a_2$  \\  
 \hline
 $\nu_e e \rightarrow \nu_e e$ & $\sin^2 \theta_{w} + 1/2$ & $\sin^2 \theta_{w}$ \\ 
 \hline
 $\bar{\nu}_e e \rightarrow \bar{\nu}_e e$ & $\sin^2 \theta_{w}$ &  $\sin^2 \theta_{w} + 1/2$  \\
 \hline
 $\nu_{\beta} e \rightarrow \nu_{\beta} e$ & $\sin^2 \theta_{w} -1/2$ & $\sin^2 \theta_{w} $  \\
 \hline
 $\bar{\nu}_{\beta} e \rightarrow \bar{\nu}_{\beta} e$ & $\sin^2 \theta_{w} $ & $\sin^2 \theta_{w} - 1/2$ \\ 
 \hline
\end{tabular}
\caption{Values of $a_1$ and $a_2$ in terms of Weinberg angle ($\theta_W$) for different flavor of neutrinos (anti-neutrinos) and $\beta$ corresponds to either $\mu$ or $\tau$.}
\label{tab2}
\end{table}

In the presence of U(1)$_X$, the $\nu -e$ scattering cross section \footnote{We have not considered the neutrino-nucleon scattering in our analysis as the constraint coming from this process is much more weaker than neutrino-electron scattering.} will be modified by the additional $t$ channel $Z'$ exchange process as

\begin{eqnarray}\nonumber
\dfrac{d \sigma  (\barparen{\nu_{\alpha}} e)}{dT}\bigg\vert_{\rm{_{_{Z'}}}}& =& \dfrac{(g_X)^4 (Q^l_X)^2 m_e}{4\pi E^2_{\nu} (2m_eT +m^2_{Z'})^2}[(2E^2_{\nu} -2E_{\nu}T + T^2)(b^2_1 + b^2_2)\pm 2b_1b_2(2E_{\nu}-T)T \\ 
&-& m_e T (b^2_1 - b^2_2)],
\label{BSM-1}
\end{eqnarray} 
where $\alpha \in (e,\mu,\tau)$. The negative sign in the last but one term corresponds to anti-neutrino.
The contribution of the new interference term to $\nu -e$ scattering induced by the $Z^\prime$ can be written as

\begin{eqnarray}\nonumber
\dfrac{d \sigma (\nu_e e)}{dT}\mid_{\rm{_{_{\rm{int}}}}} &=& \dfrac{G_F (g_X)^2 Q^l_X m_e}{\sqrt{2}\pi E^2_{\nu} (2m_eT + m^2_{Z'})}[2E^2_{\nu}(b_1+b_2) + (2E^2_{\nu} -2E_{\nu}T + T^2)(b_1c_1+ b_2c_2) \\
 &+& T(2E_{\nu} -T)(b_1c_2+b_2c_1) -m_eT(b_1-b_2 + b_1c_1 - b_2c_2) ],
\end{eqnarray} 

\begin{eqnarray}\nonumber
\dfrac{d \sigma (\bar{\nu}_e e)}{dT}\mid_{\rm{_{_{\rm{int}}}}} &=& \dfrac{G_F (g_X)^2 Q^l_X m_e}{\sqrt{2}\pi E^2_{\nu} (2m_eT + m^2_{Z'})}[2(E_{\nu}-T)^2(b_1+b_2) + (2E^2_{\nu} -2E_{\nu}T + T^2) (b_1c_1+ b_2c_2)\\
& - &T(2E_{\nu} - T)(b_1c_2+b_2c_1) -m_eT(b_1-b_2 + b_1c_1 - b_2c_2) ],
\end{eqnarray} 

\begin{eqnarray}\nonumber
\dfrac{d \sigma (\barparen{\nu_{\beta}} e)}{dT}\bigg\vert_{\rm{_{_{\rm{int}}}}} &=& \dfrac{G_F (g_X)^2 Q^l_X m_e}{\sqrt{2}\pi E^2_{\nu} (2m_eT + m^2_{Z'})}[(2E^2_{\nu} - 2E_{\nu}T + T^2)(b_1c_1 +b_2c_2) \pm T(2E_{\nu}-T)\\
&\times &(b_1c_2+b_2c_1) -m_eT(b_1c_1-b_2c_2)]
\label{BSM-2}
\end{eqnarray}
where $c_1 = -1/2 +2\sin^2\theta_W$, $c_2=-1/2$ with $\beta \in (\mu,\tau)$ and  $b_1 = \dfrac{Q^\ell_{X} + Q^{e_R}_{X}}{2}$ and $b_2 = \dfrac{Q^\ell_{X} - Q^{e_R}_{X}}{2}$ from Eq.~\ref{Lint} and Tab.~\ref{tab1} respectively. Finally combining Eqs.\ref{SM-1}-\ref{BSM-2} we find 
\begin{eqnarray}
\dfrac{d \sigma (\nu e)}{dT}= \dfrac{d \sigma (\nu e)}{dT}\bigg\vert_{\rm{_{_{SM}}}} + \dfrac{d \sigma (\nu e)}{dT}\bigg\vert_{\rm{_{Z'}}} + \dfrac{d \sigma (\nu e)}{dT}\bigg\vert_{_{\rm{int}}}. 
\label{X-sec1}
\end{eqnarray}

\begin{figure}[htp!]
\centering
\includegraphics[width=7.8cm]{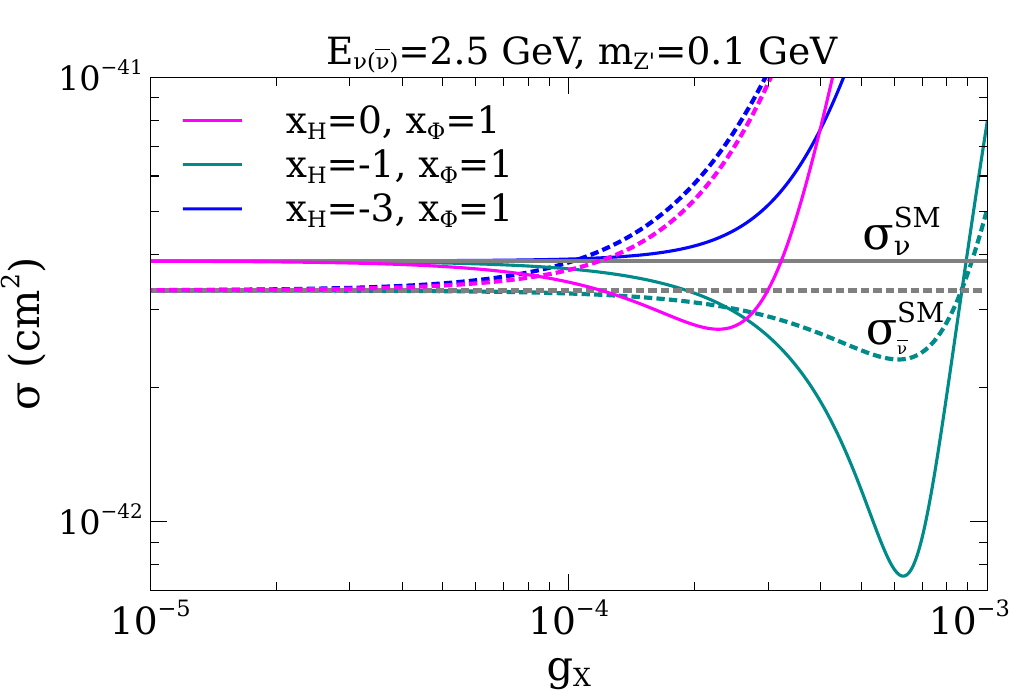}
\includegraphics[width=7.8cm]{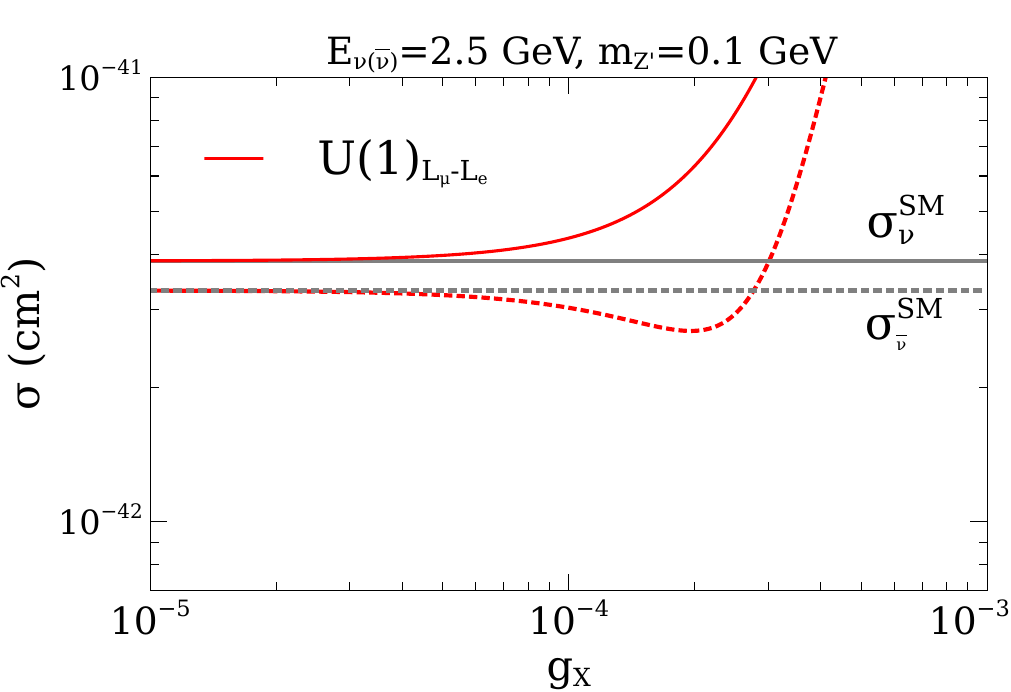}
\caption{Total cross section of $\barparen{\nu_{\mu} }-e$ scattering for SM and U$(1)_X$ scenarios. The solid (dashed) line corresponds to $\nu_{\mu}$ ($\bar{\nu}_{\mu}$) mode. See text for details.}
\label{Xsec}
\end{figure}

The interference term contributes distinctly for neutrino and anti-neutrino modes for various values of $x_H$ and $x_{\Phi}$ in U(1)$_X$ model. For example, in Fig. \ref{Xsec} we show the behavior of the cross section for muon type neutrino and anti-neutrino as a function of the gauge coupling strength $g_X$ for $x_H=0,-1,-3$ with $x_{\Phi}=1$ and U(1)$_{L_{\mu}-L_e}$. The energy of the incoming (anti) neutrino is fixed at DUNE peak energy which is nearly $2.5$ GeV. The solid and dotted lines correspond to the neutrino and anti-neutrino modes respectively. The horizontal lines represent the SM prediction of the cross section at $E_{\nu(\bar{\nu})}\simeq 2.5$ GeV. As expected, when $g_X$ is very small, both the SM and U(1)$_{X}$ values of the cross section remain almost equal. But with the increase in $g_X$, both neutrino and anti-neutrino cross section starts to deviate from the SM values. The qualitatively different behavior of the cross section of $\nu_{\mu}$ and $\bar{\nu}_{\mu}$ is clearly visible for different choices of $x_H$. The magenta lines show the variation of the cross section for $x_H=0$ and $x_{\Phi}=1$, \textit{i.e.}, U(1)$_{B-L}$ case. In this scenario, the anti-neutrino cross section (dotted magenta line) rises continuously above the SM values with the increase in $g_X$ while neutrino cross section (solid magenta line) drops below the SM prediction, attains a minimum value at $g_X \simeq 2.3\times 10^{-4}$ and then it rises very rapidly. The cross section for both SM and U(1)$_X$ becomes equal at $g_X \simeq 3\times 10^{-4} $ and we call this region as a \textit{degenerate region}. The drop in the neutrino cross section arises due to the negative contribution coming from the interference term as in Eq. \ref{BSM-2}. But the pure $Z'$ contribution is positive and it grows with $g^4_X$. At the degenerate region the contribution coming from U$(1)_X$ vanishes, \textit{i.e.}, the contributions from the interference term and pure $Z^\prime$ cancel each other. At some critical values of $g_X$, depending on $x_H$ and $x_{\Phi}$, the pure $Z'$ contribution starts to dominate and the cross section continues to rise rapidly beyond the SM prediction. For anti-neutrino, the interference term gives positive contribution to the cross section. As a result of this, the cross section rises above the SM values from the beginning and we will not get any degenerate region in this case. The behavior of the cross section changes for U(1)$_{L_{\mu}-L_e}$ compared to U$(1)_{B-L}$ as the interference term changes its sign. Here in neutrino mode the cross section rises continuously above the SM prediction while in ant-neutrino mode the cross section drops from the SM prediction and then starts increasing after crossing some critical value of $g_X$. Depending on $x_H$ and $x_{\Phi}$ values, the qualitative behavior of the cross section changes in the neutrino and anti-neutrino mode in general U$(1)_X$ scenario.
The interference term could contribute positively (or negatively) both in neutrino and anti-neutrino modes for $x_H=-3(-1)$ with $x_{\Phi}=1$. Any other combination of $x_H$ and $x_{\Phi}$ will mimic these four possibilities. Following interesting scenarios can arise in the total events rate depending on the values of $g_X$:

(i) Neutrino events will be more than the SM prediction while the anti-neutrino events will be less ($L_{\mu} -L_e$ case).

(ii) Anti-neutrino events will be more compared to SM expectation while neutrino will be less ($B-L$ case).

(iii) There is an enhancement in both neutrino and anti-neutrino events as compared to SM projection ($x_H=-3$ and $x_{\Phi}=1$ scenario).

(iv) There is an reduction in both neutrino and anti-neutrino events compared to SM values ($x_H=-1$ and $x_{\Phi}=1$ scenario).
  
\section{EXPERIMENTAL DETAILS}
\label{Exp Details}
In this section, we briefly discuss the various experiments which are relevant for our studies.\\  \\
\textbf{DUNE ND :} DUNE \cite{DUNE:2020ypp} is an upcoming super-beam long-baseline neutrino experiment. It will also have a near detector complex to measure the neutrino flux precisely. 
In our analysis we have considered a uniform beam power of 1.2 MW delivering $1.1 \times 10^{21}$ protons on target/year for the entire run of 7 years equally divided both in neutrino and anti-neutrino mode. The detector considered is a 75 ton liquid Argon near detector. This results in a total exposure of 630 MW-ton-year with 315 MW-ton-year each for neutrino and antineutrino runs.
 The detector will have an excellent energy and angular resolution for the scattered electron. The predicted fluxes for neutrino and anti-neutrino modes are taken from \cite{Marshall:2019vdy}. The small amount of contaminated $\nu_e \, (\bar{\nu}_e)$ flux could produce the background for $\nu - e$ scattering via the charged current (CC) interaction if the hadronic activity is below the detector threshold level ($\sim 50$ MeV). The misidentified $\pi^0$ could also mimic the signal produced via $\nu A \rightarrow \nu \pi^0 A$ ($A$ nucleon) if one of the photon is soft and also the hadronic activity is below the threshold.\\ \\
\textbf{TEXONO :} At the Kuo-Sheng Nuclear Power Station, the elastic $\bar{\nu}_e - e$ scattering was measured using 187 kg of CsI(Tl) scintillating crystal array with 29882/7369 kg-day of reactor ON/OFF data \cite{TEXONO:2009knm}. The neutrino and recoil electron energies vary from 3 MeV to 8 MeV. \\ \\
\textbf{CHARM II :} CHARM II experiment \cite{VILAIN1993351,VILAIN1994246} measure the electroweak parameters using the $\nu_{\mu}$ and $\bar{\nu}_{\mu}$ beam with an average energy 23.7 GeV and 19.1 GeV respectively.
The recoil electron energy range for the analysis is 3-24 GeV.

The available data of TEXONO and CHARAM II can put constraints on U(1)$_{X}$ model under consideration. To obtain the limit, we translate the bounds on U(1)$_{B-L}$ \cite{Bilmis:2015lja} to the U(1)$_X$ scenario for different $x_H$ and $x_\Phi$ by equating the cross section in both model as
\begin{equation}
(\sigma^{\rm{total}}){_{_{{\rm{U(1)}}_{B-L}}}}= \, (\sigma^{\rm{total}}){_{_{{\rm{U(1)}}_{X}}}}.
\end{equation}
\textbf{BOREXINO :} $^7$Be solar neutrino ($\nu_e$) of 862 keV energy was measured by BOREXINO collaboration \cite{Bellini:2011rx} via the neutrino electron scattering using a liquid scintillator detector. The energy range of the recoil electron is $270-665$ keV. 
Solar neutrinos ($\nu_e$) change their flavor during propagation from sun to earth.
Hence the constraint on U(1)$_{B-L}$ \cite{Bilmis:2015lja} from BOREXINO data can be translated into the constraint on $g_X$ as
\begin{equation}
\Big( \sigma_{e}  \langle P_{ee} \rangle + \sigma_{\mu / \tau} (1-\langle P_{ee} \rangle)\Big )\Big |{_{_{{\rm{U(1)}}_{B-L}}}}=\Big( \sigma_{e}  \langle P_{ee} \rangle + \sigma_{\mu / \tau} (1-\langle P_{ee} \rangle)\Big )\Big |{_{_{{\rm{U(1)}}_{X}}}}.
\end{equation}
For $L_{\mu}-L_e$ model, we get the constraint on coupling strength ($g_X$) as
\begin{equation}
\Big( \sigma_{e}  \langle P_{ee} \rangle + \sigma_{\mu / \tau} (1-\langle P_{ee} \rangle)\Big )\Big |{_{_{{\rm{U(1)}}_{B-L}}}}=\Big( \sigma_{e}  \langle P_{ee} \rangle + \sigma_{\mu} \langle P_{e\mu} \rangle\Big )\Big |{_{_{{\rm{U(1)}}_{L_{\mu}-L_e}}}} + \sigma^{\rm{SM}}_{\tau} \langle P_{e\tau} \rangle ,
\end{equation}
where $\langle P_{ee} \rangle \simeq 0.536$, $\langle P_{e\mu} \rangle \simeq 0.252$, and $\langle P_{e\tau} \rangle \simeq 0.212$ \cite{Khan:2019jvr}.\\ \\
\textbf{Electron Beam Dump :} The electron beam dump experiments provide a  significant constraint for the lower mass region of $Z'$. The constraints on the dark photon ($\gamma'$) searches at E141 \cite{Andreas:2012mt,Bjorken:2009mm} and ORSAY \cite{DAVIER1989150} can be mapped to the coupling strength ($g_X$) for various values of $x_H$ and $x_{\Phi}$. The constraint on the upper region of $\gamma'$ is approximately scaled as \cite{Ilten:2018crw}
\begin{eqnarray}
 \tau_{\gamma'}(e\epsilon^{\rm{max}}) \sim \tau_{Z'}(g_X^{\rm{max}}),
\end{eqnarray}
whereas for lower region of $\gamma'$
\begin{eqnarray}
g_X \sim \sqrt{\dfrac{(e\epsilon)^2 Br(\gamma' \rightarrow e^+ e^-)\, \tau_{Z'}}{(b^2_1+b^2_2)Br(Z' \rightarrow e^+ e^-)\, \tau_{\gamma'}}},
\end{eqnarray}
where $\epsilon$ is the kinetic mixing parameter for $\gamma'$. $\tau_{\gamma'}$ and $\tau_{Z'}$ are the lifetimes of $\gamma'$ and $Z'$ respectively. $b_1$ and $b_2$ are defined below the Eq. \ref{BSM-2}. \\ \\
\textbf{BaBaR :} BaBar \cite{BaBar:2014zli} searched for dark photon ($\gamma'$) via $e^+e^- \rightarrow \gamma' \, \gamma , \gamma' \rightarrow e^+e^-, \mu^+ \mu^-$. The new $Z'$ is also produced at BaBar via the same process and it could decay to $e^+e^-$ or $\mu^+ \mu^-$ pair. The constraint on the coupling strength is scaled as
 \begin{eqnarray}
g_X \sim \sqrt{\dfrac{(e\epsilon)^2 Br(\gamma' \rightarrow e^+ e^-/\mu^+\mu^-)}{(b^2_1+b^2_2)Br(Z' \rightarrow e^+ e^-/\mu^+\mu^-)}}.
\end{eqnarray} 
\textbf{COHERENT :} The COHERENT experiment \cite{COHERENT:2017ipa,
COHERENT:2018imc,COHERENT:2020iec} measures the neutrino-nucleus coherent elastic scattering in cesium-iodide (CsI) and argon (Ar). This can provide a test of SM as well as BSM scenarios (\textit{e.g.} a light vector mediator). In our model, the new $Z'$ couples to both neutrino and quarks with charges as described in Tab. \ref{tab1}. Here only the vector part contributes coherently as the axial vector will give the spin dependent contribution. For U(1)$_{B-L}$, the $Z'$ couples to proton (uud) and neutron (udd) with equal strength ($g_X$) and hence the cross section gets the contribution from the total number of protons and neutrons of the material. In general for U(1)$_X$, the $Z'$ couples to proton and neutron with different strength. For instance, $x_H=-3, x_{\Phi}=1$, the $Z'$ coupling to proton (neutron) is -1.25~$g_X$ (0.25~$g_X$). In this case, there is a mutual cancellation among the proton and neutron contributions. Thus in this scenario, the COHERENT experiment can not provide as tighter a constraint as the U(1)$_{B-L}$ model. For the $L_{\mu}-L_e$ case, $Z'$ couples to quarks via loop and the constraint is an order of magnitude weaker than other scenarios as shown in Fig. \ref{chi2_DUNE_com}. To derive the bound, we follow the method as described in \cite{Cadeddu:2020nbr}.

\section{Results}
\label{results_DUNE}
At DUNE ND, the $\nu-e$ scattering events are calculated by
\begin{eqnarray}
N_{\rm{events}} = \int dE_{\nu}\, dT \, \dfrac{d\Phi}{dE_{\nu}} \, \dfrac{d\sigma}{dT} \, \eta
\end{eqnarray}
where  $\dfrac{d\Phi}{dE_{\nu}}$ is the incoming neutrino flux \cite{Marshall:2019vdy} at the detector and $\eta$ is the efficiency to detect an electron in the final state. To quantify the effect of U(1)$_X$, we perform $\chi^2$ analysis in two different ways - (i) using total number of events; (ii) bin by bin analysis. 

\subsection{ Rate only analysis}

In this case, $\chi^2$ is defined as 
\begin{eqnarray}
\chi^2 ={\rm{min}}\bigg [\dfrac{(N^{\rm{tot}}_{_{\rm{NP}}} - (1+\alpha)N_{_{\rm{SM}}} - (1+\beta)N_{_{\rm{BG}}})^2}{N^{\rm{tot}}_{_{\rm{NP}}}} + \dfrac{\alpha^2}{\sigma^2} + \dfrac{\beta^2}{\sigma^2}\bigg],
\end{eqnarray}
where $N_{_{\rm{SM}}}$ and $N_{_{\rm{BG}}}$ are the total number of events for SM signal and background respectively. $N^{\rm{tot}}_{_{\rm{NP}}}$ represents the total number of events in the presence of new physics scenario under consideration including the background. We use the estimated background corresponding to the charged current quasi elastic scattering and misidentified $\pi^0$ events as given in \cite{deGouvea:2019wav}. $\alpha$ and $\beta$ are two nuisance parameters with mean value at zero and $\sigma$ is equal to $5\%$ systematic uncertainties. We take the minimum value of $\chi^2$ after varying over $\alpha$ and $\beta$. In our analysis, we consider $\eta$ to be 0.95 to match the event distribution in reference \cite{deGouvea:2019wav}.

\begin{figure}[htp!]
\centering
\includegraphics[width=7.7cm]{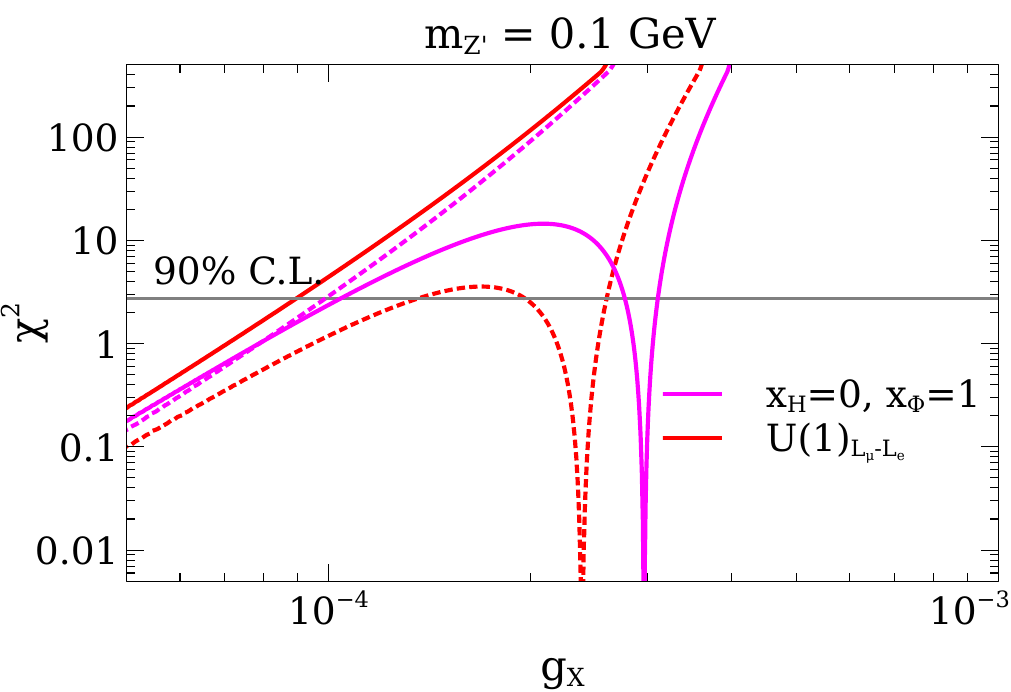}
\includegraphics[width=7.7cm]{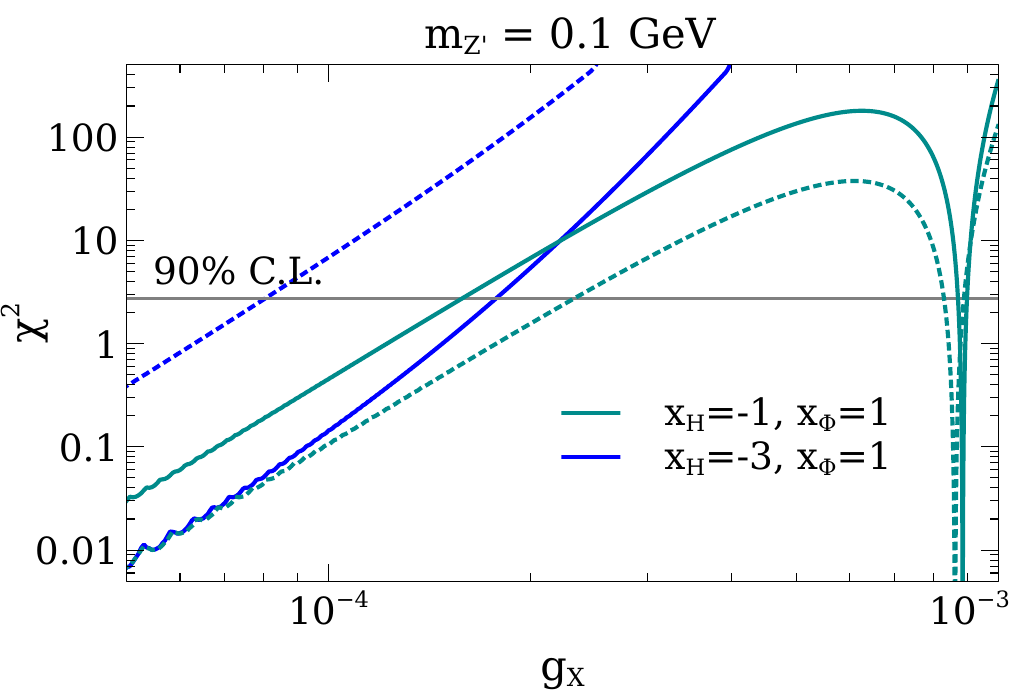}
\caption{$\chi^2$ as a function of the coupling constant for neutrino (solid) and anti-neutrino (dashed) modes using the total rate at DUNE ND.}
\label{chi2BL}
\end{figure}

In Fig. \ref{chi2BL}, we show the $\chi^2$ as a function of the gauge coupling strength $g_X$ for representative values of $x_H=0,-1,-3$ with $x_{\Phi}=1$ and U$(1)_{L_{\mu}-L_e}$ scenarios. The solid (dashed) line represents the $\chi^2$ for neutrino (anti-neutrino) mode. The left panel in Fig. \ref{chi2BL} shows the $\chi^2$ for $x_H=0$, $x_{\Phi} =1$ ( \textit{i.e.} U$(1)_{B-L}$) and  U$(1)_{L_{\mu}-L_e}$ scenario. It is apparent from the plot that
for $x_H=0$ and $x_{\Phi}=1$, the $\chi^2$ rises continuously as  $g_X$ increases for anti-neutrino mode as shown by the dotted magenta line.  In this case, the value of $g_X (>9\times10^{-5})$ is seen to be ruled out by DUNE ND at 90 $\%$ C.L for $m_{Z^\prime}=0.1$ GeV. For the neutrino mode and $x_H=0$ and $x_{\Phi}=1$, a sharp decline is observed in the constraint plot. This feature arises due to the negative contribution coming from the interference terms as shown in Fig. \ref{Xsec}. The $\chi^2$ vanishes near the degenerate region when the SM and U$(1)_X$ cross section becomes equal. The negative contribution of the interference term actually reduces the capability of the neutrino mode near the degenerate region to constrain the U$(1)_X$ scenario for $x_H=0$ and $x_{\Phi}=1$. This difficulty can be overcome by performing a bin by bin analysis as will be shown later. Note that for the $L_{\mu}-L_e$ scenario, the neutrino and antineutrino modes demonstrate an opposite behavior as compared to the B-L scenario. The right panel in Fig. \ref{chi2BL} shows the $\chi^2$ for two different sets of illustrative values of $(x_H,x_\Phi)$. For $x_H = -3$ and  $x_\Phi = 1$, there is no degenerate region and the $\chi^2$ increases with increasing values of $g_X$ for both neutrino and antineutrino channels. The antineutrino contribution is seen to be significantly higher than the neutrino mode since the interference term reinforces the cross-section. On the other hand for $x_H = -1$ and  $x_\Phi = 1$ both neutrino and antineutrino $\chi^2$ depict a sharp drop corresponding to the degenerate region.

\begin{figure}[h]
\centering
\includegraphics[width=8cm]{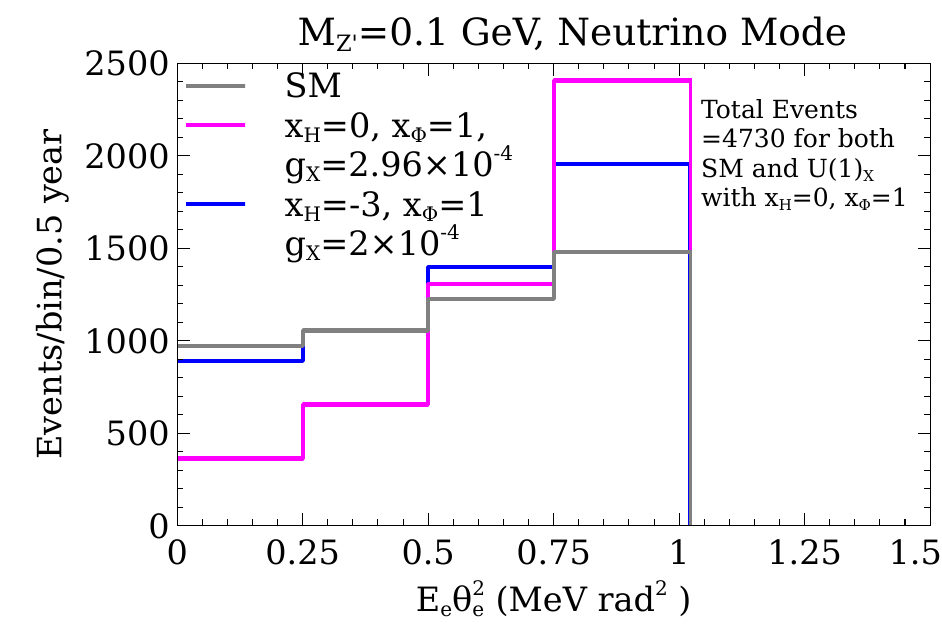}
\includegraphics[width=8cm]{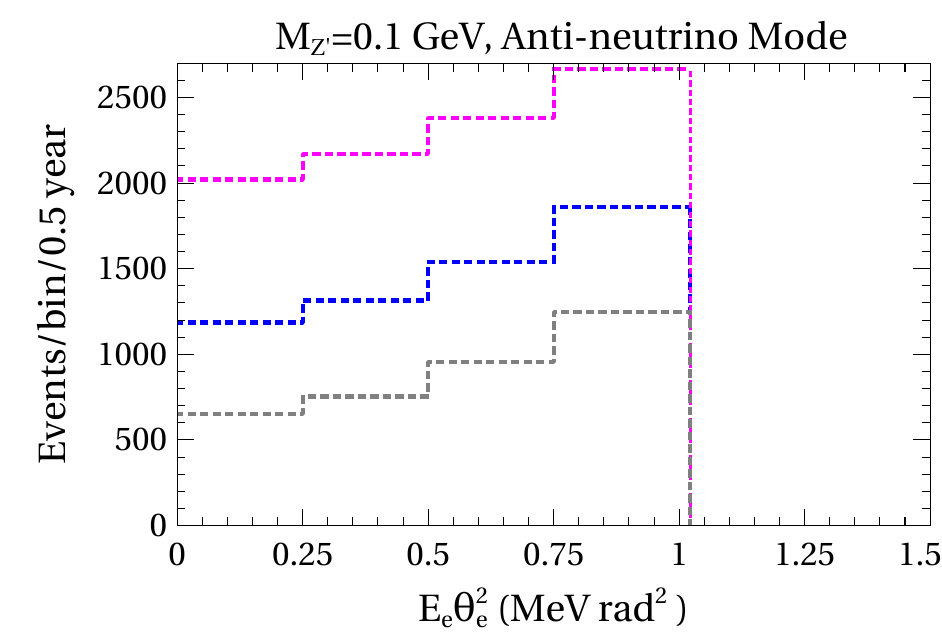}
\caption{Neutrino (anti-neutrino) events for the SM and U(1)$_{\rm{X}}$ scenarios.}
\label{Events-B-L}
\end{figure} 

\subsection{Bin by bin analysis}
In this section, we present the results for bin by bin analysis of $\nu-e$ scattering at DUNE ND. 
The binning is done in terms of the kinematic variables $E_e \theta^2_e$ \cite{deGouvea:2019wav}. Here $E_e (= T+m_e)$ and $\theta_e$ are the total energy of the scattered electron and the angle between the scattered electron and the beam direction respectively. We consider the energy of the scattered electron 
to be $0.05\,\, {\rm MeV} \,<E_e <15$ GeV. We employ the kinematic cuts $E_e \theta^2_e<2m_e$ which help to reduce the background events in the analysis.
In Fig. \ref{Events-B-L}, we depict the number of neutrino and anti-neutrino events at DUNE ND as a function of $E_e \theta^2_e$ bins. Note that we have neglected the effect of energy and angular resolution in Fig. \ref{Events-B-L}. It was shown in reference \cite{deGouvea:2019wav}, that the energy resolution does not play a very significant role in changing the event distribution. But the angular resolution, can affect the spectrum. In our later analysis we have included the effect of angular resolution. Though the $\nu-e$ scattering cross section is small, the total number of events is large due to the high intensity flux at DUNE ND. The left panel in Fig. \ref{Events-B-L} is for the neutrino mode.  The magenta and blue lines correspond to different U$(1)_X$ scenarios while the gray line is for the SM case. The magenta line corresponds to the U$(1)_{B-L}$ scenario and for this we choose the value of $g_X$ such that we encounter   the degenerate  region for the neutrino mode leading to the same total  number of events as SM. 
This corresponds to the destructive interference effect  as discussed earlier. However even though the total number of events are same, the distribution of events are significantly different in each bin. The number of events decreases from the SM values in the first two bins while it increases above the SM in the last two bins. This indicates that if a bin by bin analysis is performed then the  effect of destructive interference in reducing the sensitivity can be tackled. The blue line corresponds to $x_H=-3$ and $x_\Phi=1$, the $g_X$ chosen for this is such that the neutrino cross-section starts departing from the SM value near DUNE peak energy as can be seen from Fig. \ref{Xsec}. 
For anti-neutrino mode, the number of events increases above the SM values for all the bins. 

\begin{figure}[h]
\centering
\includegraphics[width=8cm]{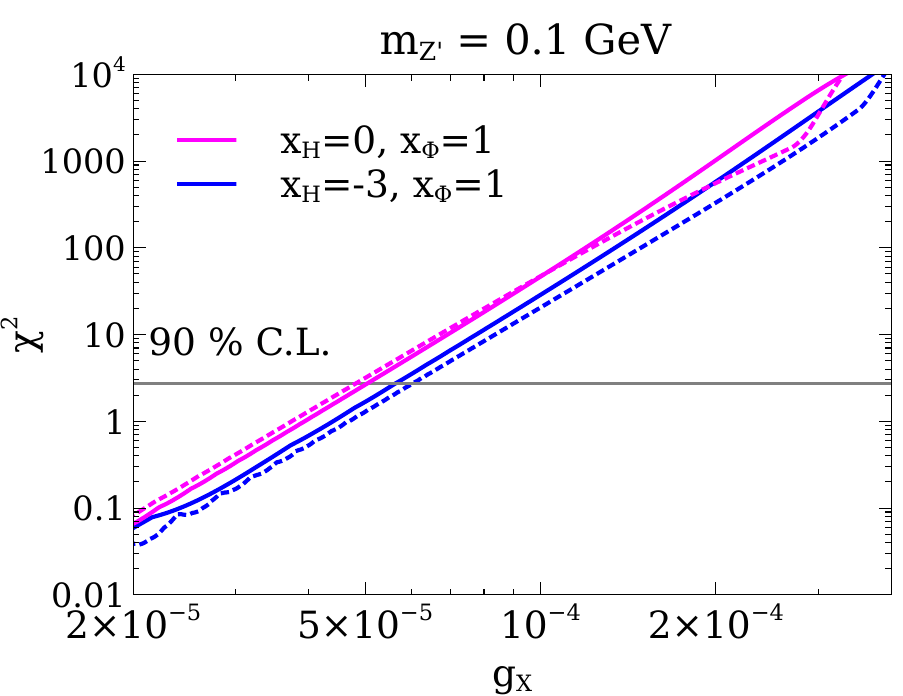}
\caption{$\chi^2$ as a function of the coupling strength $g_X$ for neutrino (solid line) and anti-neutrino (dotted line) modes using the binned spectrum.}
\label{ch2-binwise}
\end{figure}

We perform the $\chi^2$ analysis over the $E_{e} \theta^2_e$ bins as
\begin{eqnarray}
\chi^2 ={\rm{min}}\Bigg[ \sum_{i=1}^4 \dfrac{(N^{i}_{_{\rm{NP}}} - (1+\alpha)N^i_{_{\rm{SM}}} - (1+\beta)N^i_{_{\rm{BG}}})^2}{N^{i}_{_{\rm{NP}}}} + \dfrac{\alpha^2}{\sigma^2} + \dfrac{\beta^2}{\sigma^2}\Bigg],
\label{Eq-chi2-bin}
\end{eqnarray}
where $N^i_{_{\rm{SM}}}$ and $N^i_{_{\rm{BG}}}$ are the number of events for SM signal and background respectively in the i-th bin. $N^{i}_{_{\rm{NP}}}$ is the combined number of events with the U$(1)_X$ and background in the i-th bin. In Fig. \ref{ch2-binwise}, we show the $\chi^2$ performed over $E_{e} \theta^2_e$ bins as in Eq. \ref{Eq-chi2-bin} for two sets of values of $x_H$ and $x_{\Phi}$. Now there is no sharp decline like behavior present in the neutrino mode for $x_H=0$ and $x_{\Phi}=1$ as the events in each bin differ from the SM prediction though the total events are equal as shown in Fig. \ref{Events-B-L}. Hence, the effect of the interference terms will not matter much if the analysis is performed over $E_{e} \theta^2_e$ bins. From Fig. \ref{ch2-binwise} we find that $g_X \gtrsim 4.5 \times 10^{-5}$ is ruled out as opposed to the $g_X \gtrsim 9\times 10^{-5}$ obtained in the rate only analysis for $x_H=0$ and $x_{\Phi}=1$. Both the neutrino and anti-neutrino modes provide almost equal bounds on $g_X$ compared to the total rate analysis shown in Fig. \ref{chi2BL}. Thus the bin by bin analysis results in a twofold improvements in the overall bounds. 

\begin{figure}[h]
\centering
\includegraphics[width=8cm]{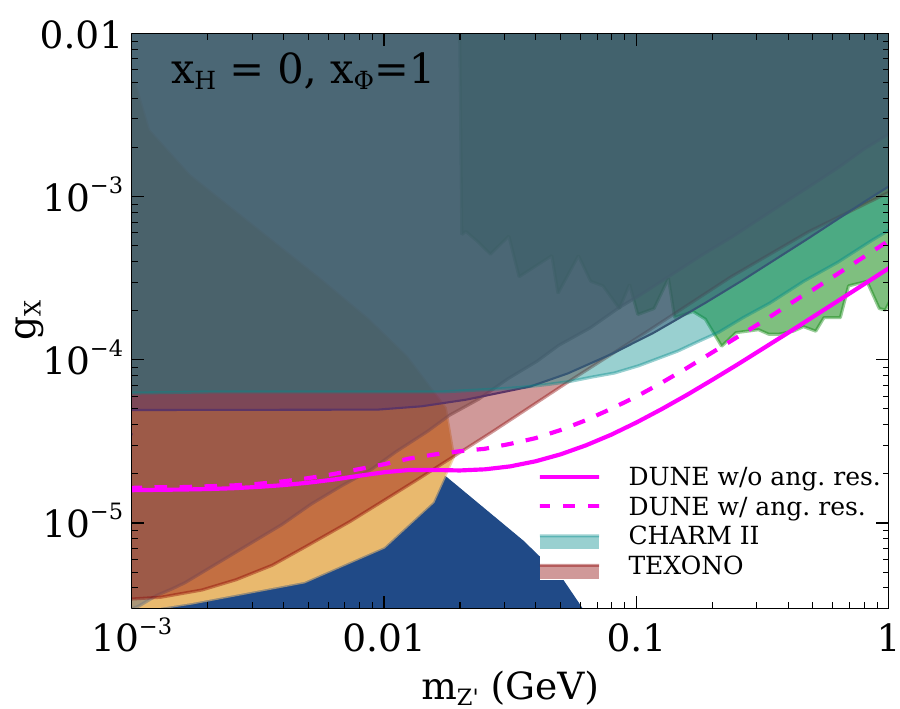}
\includegraphics[width=8cm]{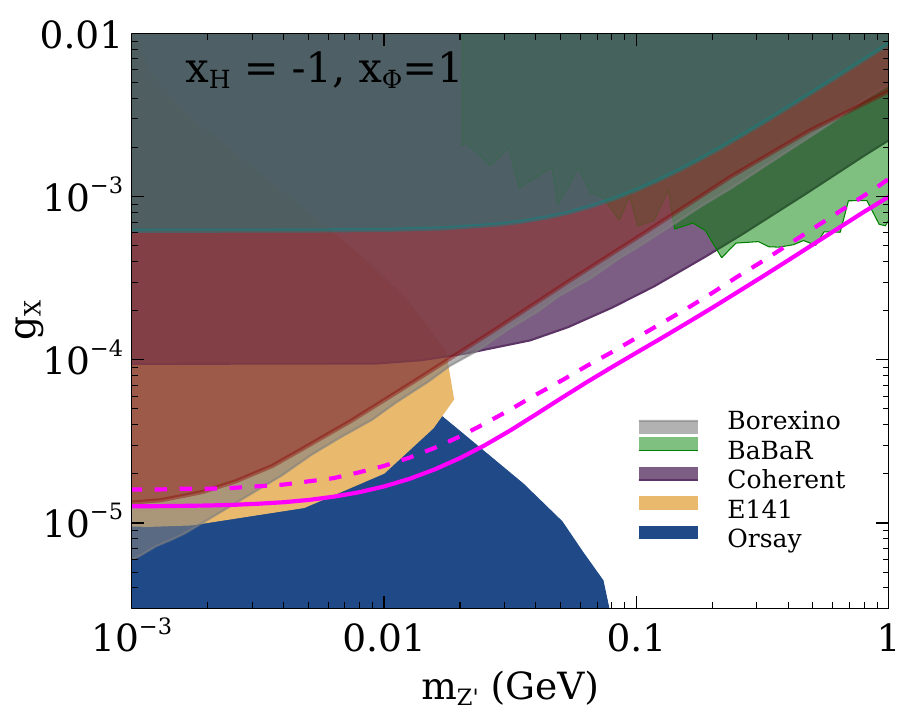}
\includegraphics[width=8cm]{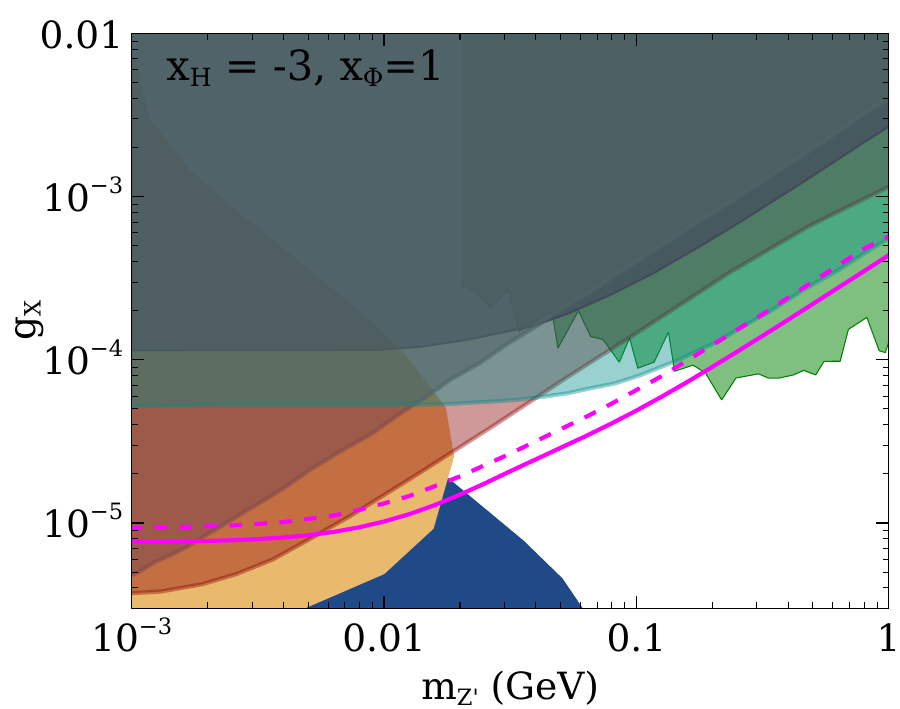}
\includegraphics[width=8cm]{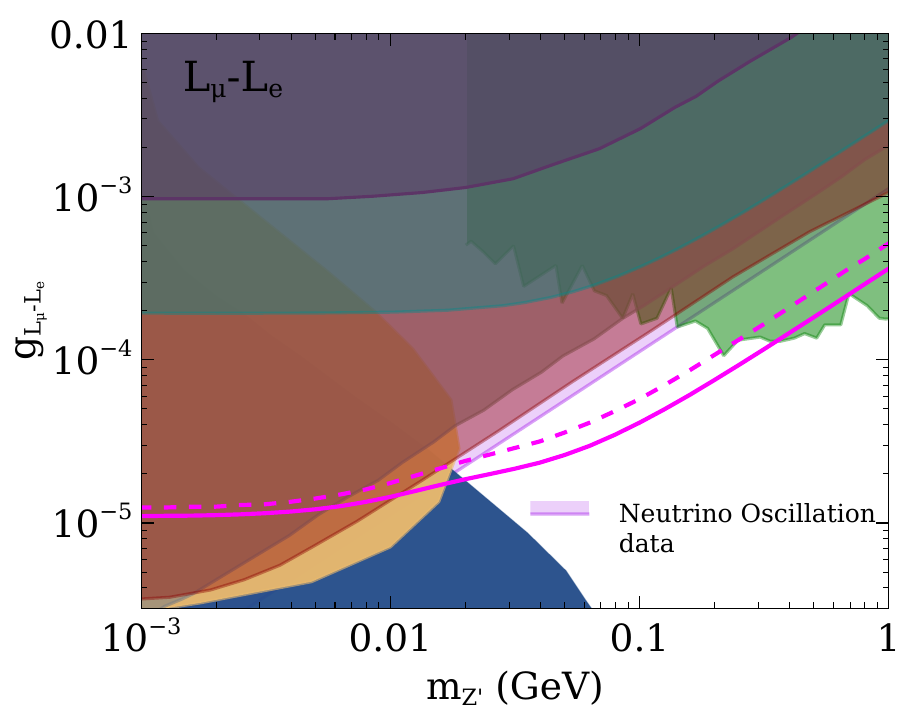}
\caption{90$\%$ CL contour on the $g_X$-m$_{z'}$ plane.}
\label{chi2_DUNE_com}
\end{figure}

Our main results are shown in Fig. \ref{chi2_DUNE_com} where we depict the constraints coming from different experiments on the $g_X - m_{Z'}$ plane for representative values of $x_H$ and $x_{\Phi}$. The magenta lines are for DUNE ND with 90$\%$ C.L. For obtaining these bounds, we
define the total $\chi^2$ as
\begin{eqnarray}
\chi^2_{_{\rm{tot}}} = \chi^2_{\nu} + \chi^2_{\bar{\nu}}
\label{tot_chi}
\end{eqnarray}
\textit{i.e.} we combine the neutrino and ant-neutrino mode using Eq. \ref{Eq-chi2-bin} and treating the systematic uncertainties independently for both the modes. We also show the effect of including the angular resolution ($\sigma_{\theta}=1^{\circ}$). The solid magenta line corresponds to the scenario without any angular resolution function while the dotted magenta line shows the effect of angular resolution function. In the presence of the angular resolution function, the sensitivity deteriorates slightly.

We also show the constraints coming from the electron beam dump experiments like E141 (brown shaded region) and Orsay (blue shaded region) in Fig.~\ref{chi2_DUNE_com}. The beam dump experiments are seen to constrain the region with 
lighter $Z^\prime$, for example $ m_{Z^\prime} \lesssim 15$ MeV.  The constraints coming from BABAR are shown by the green shaded region. This puts constraints on the heavier $Z^\prime$, for example $m_{Z'} \gtrsim 300$ MeV and $g_X > 10^{-4}$ depending on the choices of $x_H$ and $x_{\Phi}$. The $\nu-e$ scattering experiments such as Borexino (grey shaded), TEXONO (red shaded), and CHARM II (cyan shaded) can put significant constraint on $g_X - m_{Z^\prime}$ plane covering the full range of $m_{Z^\prime}$ presented in the figure. Neutrino oscillation data puts bound on the flavor non-universal $L_{\mu}-L_e$ model \cite{Coloma:2020gfv} as shown by the violet shaded region while the constraint coming from the COHERENT experiment is depicted by purple shaded region.

It is seen from Fig. \ref{chi2_DUNE_com} that DUNE can probe parameter spaces not accessible by the other experiments and can improve the bound for certain ranges of the $m_{Z^\prime}$ depending on the model under consideration. 
The best constraint comes for $x_H=-1$ and $x_{\Phi}=1$ case where we obtain significant improvement as compared to the present constraint in the range of 20 MeV $<m_{Z^\prime}<300$ MeV  from DUNE ND. 
For the U(1)$_{B-L}$ case, DUNE can improve the constraints coming from other experiments in the range of 15 MeV $<m_{Z'}<200$ MeV as seen from the top left panel. 

Note that the analysis in reference \cite{Dev:2021xzd} did not report this improvement in their combined neutrino and anti-neutrino runs. The main reason for that is the authors have considered $\chi^2_{\nu + \bar{\nu}}$ (i.e. combined at events level) whereas we have added the $\chi^2$s separately as in Eq. \ref{tot_chi} since the neutrino and antineutrinos are coming from different runs. Moreover, we have performed a bin by bin analysis which helps in ameliorating the effect of destructive interference. 

\section{Conclusions}
\label{Conclusion}
In this work, we show the capability of the proposed DUNE ND to constrain a general U$(1)$ model. The hallmark of such models is an extra neutral gauge boson ($Z^\prime$) and a singlet Higgs. The U$(1)$ charges of the fermions can be expressed in terms of those of the scalars from the anomalies cancellation conditions. We focus on the possibility of constraining the interaction strength and mass of this extra $Z^\prime$ using $\nu-e$ scattering at DUNE ND. The presence of $Z^\prime$ can give rise to interference effects in such a process. Depending on the U$(1)$  scenario four typical situations can arise: (i) destructive interference in only the neutrino channel, 
(ii) destructive interference in only the antineutrino channel, (iii) destructive interference in both neutrino and anti-neutrino channel, and (iv) no destructive interference in either channel. 
Note that, for  U$(1)_{B-L}$ which is a  popular special case of a general  U$(1)$  model  we have the scenario (i) whereas, for U$(1)_{L_\mu- L_e}$ case one gets the scenario (ii). However, within the ambit of general U$(1)$ models, two more cases can arise which are pointed out in our work. Depending on the scenario chosen one can get more number of either neutrinos or antineutrinos or a reduction or enhancement  in both as compared to SM expectations depending on the values of $g_X$. 
We prescribe a bin by bin analysis and point out the salient features of  this in comparison to analysis considering total rates using only neutrino and antineutrino runs. 
We show that the effect of destructive interference which spoils the sensitivity of either the neutrino or the antineutrino mode or both depending on the U$(1)_X$ charges, can be overcome using a bin by bin analysis. Therefore such an analysis can take advantage of the combined statistics of neutrino and antineutrino modes to improve on the results. 
In such an analysis, both neutrino and antineutrino mode gives similar contribution for U$(1)_{B-L}$ scenario even though the neutrino channel is affected by the interference effect. 
Consequently, there is a twofold improvement in the bounds on $g_X$ when we perform a bin by bin analysis using both neutrino and antineutrino channels. Finally, we present the constraints on the $g_X - m_{Z^\prime}$ plane for four different cases outlined above. We also compare our results with that obtained from other electron scattering experiments like TEXONO, Borexino, CHARM-II, COHERENT, beam dump experiments, and Babar respectively. We show that the electron scattering measurements at DUNE ND can probe areas which were hitherto unconstrained by any other experiments, thus providing the strongest bounds so far.

\textbf{ACKNOWLEDGEMENTS}\\ \\
We  would like to thank Pedro A.N. Machado, Roberto Petti, Jaydip Singh, and  Tanmay Kumar Poddar for useful discussions. We also thank Pilar Coloma for useful suggestion and providing the data for $L_{\mu}-L_e$ model. S.G. acknowledges the J.C Bose Fellowship (JCB/2020/000011) of Science and Engineering Research Board of Department of Science and Technology, Government of India.

\bibliographystyle{apsrev}
\bibliography{NU-bib.bib}
\end{document}